\documentclass[sn-aps]{sn-jnl}

\jyear{2021}%

\theoremstyle{thmstyleone}

\theoremstyle{thmstyletwo}%

\theoremstyle{thmstylethree}%

\newcommand{\n}{{(n)}} 
\newcommand{\mE}{\mathfrak{E}}

\renewcommand{\AA}{\mathcal{A}} 
\newcommand{\PP}{\mathcal{P}}

\newcommand{\eea}{\end{eqnarray*}} 
\renewcommand{\Re}{\text{Re }}

\newcommand{\Lra}{\Leftrightarrow}

\newcommand{\es}{\emptyset} 
\newcommand{\app}{\approx} 
 
\newcommand{\Ra}{\Rightarrow}

\newcommand{\XX}{\mathcal{X}}

\newcommand{\tit}{\textit} 
 
\newcommand{\ity}{\infty} 
 
\newcommand{\mo}{{-1}} 
\newcommand{\bg}{\begin} 
 
\newcommand{\en}{\end} 
 
\newcommand{\tx}{\text}

\newcommand{\De}{\Delta} 
\newcommand{\si}{\sigma}

\newcommand{\lam}{\lambda} 
 
\newcommand{\HH}{\mathcal{H}} 
\newcommand{\om}{\omega} 
\newcommand{\dg}{\dagger} 
\newcommand{\se}{\subseteq} 

\renewcommand{\R}{\mathbb{R}}

\newcommand{\Om}{\Omega}

\renewcommand{\SS}{\mathcal{S}}

\newcommand{\KK}{\mathcal{K}} 
\newcommand{\BB}{\mathcal{B}} 

\newcommand{\X}{\mathbb{X}} 
\newcommand{\K}{\mathbb{K}} 
\newcommand{\ep}{\epsilon}

\newcommand{\la}{\langle} 
\newcommand{\de}{\delta} 
\newcommand{\ra}{\rangle} 

\newcommand{\EE}{\mathcal{E}} 
\newcommand{\sa}{$\sigma$-algebra }

\newcommand{\1}{\mathbf{1}}
\newcommand{\tpsi}{\tilde \psi}

\begin{document}

\title[Nonprobabilistic typicality with application to quantum mechanics]{Nonprobabilistic typicality with application to quantum mechanics}


\author*{\fnm{Galvan} \sur{Bruno}}\email{b.galvan@virgilio.it}



\affil{\orgaddress{\street{via Milano 60}, \city{Trento}, \postcode{38122}, \country{Italy}}}




\abstract{In this paper two hypotheses are developed. The first hypothesis is the existence of random phenomena/experiments in which the events cannot generally be assigned a definite probability but that nevertheless admit a class of nearly certain events. These experiments are referred to as \textit{typicalistic} (instead of probabilistic) experiments. As probabilistic experiments are represented by probability spaces, typicalistic experiments can be represented by \textit{typicality spaces}, where a typicality space is basically a probability space in which the probability measure has been replaced by a much less structured typicality measure $T$. The condition $T(A) \approx 1$ defines the typical sets, and a typicality space is related to a typicalistic experiment by associating the typical sets of the former with the nearly certain events of the latter. Some elements of a theory of typicality, including the definition of typicality spaces, are developed in the first part of the paper.

The second hypothesis is that the evolution of a quantum particle (or of a system of quantum particles) can be considered a typicalistic phenomenon, so that it can be represented by the combination of typicality theory and quantum mechanics. The result is a novel formulation of quantum mechanics that does not present the measurement problem and that could be a viable alternative to Bohmian mechanics. This subject is developed in the second part of the paper.
}




\maketitle

\section{Introduction}

In this paper two hypotheses are developed. The first hypothesis is that there are random phenomena in nature that are nonprobabilistic, i.e., that cannot be represented by a probability space. This hypothesis is supported by the following quote of Kolmogorov~\cite{kol1}:
\begin{quote}
In everyday language we call random these phenomena where we cannot find a regularity allowing us to predict precisely their results. Generally speaking there is no ground to believe that a random phenomenon should posses any definite probability. Therefore, we should have distinguished between randomness proper (as absence of any regularity) and stochastic randomness (which is the subject of probability theory).
\end{quote}
Generalizing beyond Kolmogorov it is reasonable to hypothesize the existence of phenomena that exhibit a kind of randomness that is intermediate between the two extremes mentioned by Kolmogorov, namely, stochastic (or probabilistic) randomness and the absence of any regularity. For example, T. Fine et al. studied the randomness associated with a set of probability measures \cite{fine1,fine2,fine3}.

In this paper we consider a kind of randomness in which the only statistical regularity is the existence of a class of \tit{nearly certain} events\footnote{In this paper the adverb ``nearly'' systematically replaces the more common adverb ``almost'' in all the expressions of the type ``almost certain'', ``almost always'', and so on. The reason is that these expressions have become part of the standard terminology of probability theory with a mathematical meaning that is different from the empirical meaning of everyday language we want to assign them.}. As probabilistic experiments are represented by probability spaces, the idea is to introduce a novel kind of space, namely, a \tit{typicality space}, for representing experiments with this novel kind of randomness. Roughly speaking, a typicality space is a probability space in which the probability measure has been replaced by a less structured set function $T$ that defines the typical sets in the same way as a probability measure does, namely, by the condition $T(A) \app 1$. Such a set function is referred to as a \tit{typicality measure}. Analogous to a probability space, a typicality space is related to the experiment it represents by associating the typical sets of the space with the nearly certain events of the experiment, as better explained later. As the attribute \tit{probabilistic} is associated with usual randomness, this new kind of randomness is referred to as \tit{typicalistic} randomness. Some elements of a mathematical theory of typicality, together with the rules for relating it to the empirical world of experiments, are developed in Section \ref{typran}. The study of typicalistic randomness requires an accurate  preliminary formulation of probabilistic randomness, which is the subject of Section \ref{probran}.

\begin{figure}
\bg{center}
\includegraphics{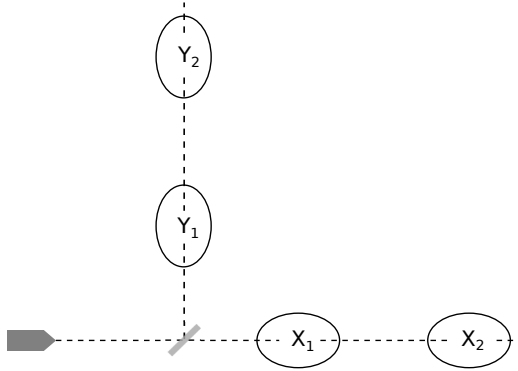}%
\end{center}
\caption{A quantum particle is emitted by a source and then split by a beam splitter. Physical intuition suggests that if the particle is detected in region $X_2$ (resp. $Y_2$) at a suitable time $t_2$, then it was in region $X_1$ (resp. $Y_1)$ at a suitable previous time $t_1$.  \label{fig01}}
\end{figure}

The second hypothesis of this paper, which motivates the first, is that the (non measured) evolution of a quantum particle (or of a system of quantum particles) can be considered as a typicalistic phenomenon. To understand this, we note first that since the position operators of the particle at two different times do not commute, the standard quantum formalism does not provide a joint distribution for the two variables, i.e., it does not provide a probability measure giving the probability that the particle is in a region $X_1$ at a time $t_1$ and in a region $X_2$ at a time $t_2$. As explained in Section \ref{quaevo}, this fact may have led some of the founding fathers of quantum mechanics to renounce the idea that a quantum particle follows a definite trajectory. On the other hand, in some situations, our physical intuition strongly suggests that the positions of the particle at two different times are correlated. For example, if a quantum particle crosses a beam splitter and is subsequently detected in region $X_2$ (resp. $Y_2$) at a suitable time $t_2$, the physical intuition suggests that the particle was in region $X_1$ (resp. $Y_1$) at a suitable previous time $t_1$ (see Fig. \ref{fig01}). The proposal is then to maintain the assumption that the particle follows a definite trajectory but to hypothesize that its evolution is a typicalistic phenomenon, and therefore, it must be represented by a typicality measure rather than by a probability measure.

In this example this measure should work as follows. Let the set of all the possible trajectories of the particle be the sample space of this experiment. Then, let $\la t, X \ra$ denote the set of all the trajectories belonging to region $X$ at time $t$, so that $\la t, X \ra$ represents the event ``the particle is in region $X$ at time $t$''. Then, the appropriate typicality measure for this experiment must define the set $\la t_1,  X_1 \ra \cap \la t_2, X_2 \ra \cup \la t_1,  Y_1 \ra \cap \la t_2, Y_2 \ra$ as \tit{typical}, so that the corresponding event is nearly certain.

The typicality measure associated with the evolution of a system of quantum particles is referred to as the \tit{quantum typicality measure}. Its physical motivation and its mathematical definition are the subjects of Sections \ref{quaevo} and \ref{typspa}, respectively. In the concluding Section \ref{disc}, the novel formulation of quantum mechanics derived from this approach is discussed and compared with Bohmian mechanics.

Let us briefly review the relevant literature about the subjects presented in this paper. The studies \cite{fine1,fine2,fine3} of T. Fine et al. about the randomness associated with sets of probability measures have already been mentioned above. The association between typical sets and nearly certain events is usually referred to as Cournot's principle; a historical presentation of this principle, with many references to related works, can be found in \cite{shafer1}. See also section \ref{frocou} for more about Cournot's principle. The notion of typicality, its relation to probability, and its role in statistical and Bohmian mechanics has been discussed in various papers and books. For example, \cite{qequilibrium, seldonboltz, volchan2007probability, goldstein2012typicality, hemmo2015probability, allori2020statistical, durr2021typicality}. The possibility of defining a nonadditive measure of typicality has been considered in \cite{seldonboltz, volchan2007probability, goldstein2012typicality}.  To the best of our knowledge, the only attempts to represent quantum evolution by means of a typicality measure are the two previous papers by this the author on this subject \cite{galvan:1, galvan:4}. The present paper is however self-contained with many improvements and one important difference with respect to the previous two that is emphasized in remark \ref{rempre}.

\section{Probabilistic randomness \label{probran}} 

To study typicalistic randomness, it is useful first to pave the way by introducing some notions about probabilistic randomness, in particular: (1) When can we say that an experiment exhibits probabilistic randomness, and (2) how is an experiment of this kind related to a probability space? 

There are various approaches to this problem in the literature. The one presented here, although inspired by Cournot's principle, is appropriate for the subsequent study of typicalistic randomness and therefore includes some definitions and propositions that, to the best of our knowledge, are novel in the literature. For this reason, in the next subsection, a short historical introduction to Cournot's principle and some logical steps leading to this particular formulation of probabilistic randomness are presented.

Two preliminary remarks are in order. The first is about terminology. In this paper, particular attention is given to distinguishing between mathematical and empirical entities or properties. Therefore, for example, a \tit{subset} of the sample space of a probability space is a mathematical entity, while an \tit{event} of an experiment is an empirical entity; these two entities are identified only if a suitable probability space is associated with a suitable experiment. Analogously, \tit{typicality} is a mathematical property of subsets of the sample space, namely, the property defined by the mathematical (vague) condition $P(A) \app 1$, while \tit{nearly certainty} is an empirical property of events, and again, these two properties may be identified only when a probability space and an experiment are associated.

Second remark. In what follows, the mentioned mathematical and empirical entities or properties are defined and related to each other, and some consequences are deduced. For clarity and simplicity, these consequences are expressed in terms of propositions and lemmas. However, the presence of empirical entities and the vagueness of some definitions makes unavoidable that these propositions and lemmas do not have the same level of rigor as the usual propositions and lemmas only involving mathematical entities and exact definitions.

\subsection{From Cournot's principle to probabilistic models \label{frocou}}

Cournot's principle states that an event with probability very close to $1$ is nearly certain (or equivalently, that an event with probability very close to $0$ is nearly impossible). It was originally introduced to provide an empirical meaning to the classical definition of probability, according to which the probability of an event is the ratio of the number of the outcomes that favour the event to the total number of outcomes. It owes its name to the fact that Augustin Cournot seems to have been the first to say explicitly that probability theory does gain empirical meaning only by declaring events of vanishing small probability to be impossible \cite{shafer1}.

The introduction of probability spaces by Kolmogorov \cite{kolmogorov1956foundations} determined the complete separation between the mathematical structure of probability and the empirical world of experiments. In other words, an experiment is no longer implicitly equipped with a probability as in the classical approach, but we now have two well-separated entities, a probability space (of mathematical nature) and an experiment (of empirical nature) that must be related. Kolmogorov's proposal for relating them (Chapter I, \textsection 2 of \cite{kolmogorov1956foundations}) in quick synthesis and partially reformulated is the following: Given a random experiment, under certain conditions (not specified by Kolmogorov) \tit{there exists} a probability space $(\Om, \AA, P)$ such that  (a) nearly certainly the relative frequency of an event $A$ in a long sequence of trials is close to $P(A)$, and (b) if $P(A) \app 1$ for some event $A$ then nearly certainly $A$ occurs in a single trial of the experiment.

We note therefore that, already in Kolmogorov's approach, Cournot's principle (point (b)) appears, rather than a principle, as a criterion for associating a probability space with an experiment. One often reads in the literature that point (a) can be deduced from point (b) by means of Bernoulli's theorem. This is correct, but in this case, point (b) is applied to the experiment composed of a large number of independent trials of the original experiment. In other words, point (a) can be replaced by the claim that point (b) applies to the experiment composed of a generic number $n$ of trials of the original experiment, and then point (a) can be derived as a pure mathematical deduction. This is the approach adopted here.

In the present paper, Kolmogorov's approach is then reformulated as follows: (1) A probability space is said to be a \tit{probabilistic model} of an experiment if point (b) is satisfied by the $n$-fold product of the probability space and the experiment composed of $n$ trials of the original experiment, for all $n \geq 1$, and (2) an experiment is said to be a \tit{probabilistic experiment} if it admits a probabilistic model. By this approach, we avoid mentioning unspecified conditions, as Kolmogorov does. Moreover, as we'll see, other than the well-known correspondence between probability and relative frequency we can also deduce the fact that an experiment admits at most one probabilistic model.

\subsection{Nearly certain events}

In this presentation we assume as already defined the following notions: on the mathematical side, the notion of probability space, and on the empirical side, the notions of experiment, outcome, sample space, event, and trial, as they are usually defined in probability theory. There are, however, two properties of the events that are usually implicitly assumed but that are not required here, namely, (i) that the relative frequencies of the events are stable in the long run and (ii) that one can always empirically determine if an event occurs or not.

There is, on the contrary, an important empirical notion for which a standard definition seems to be lacking, namely, the notion of nearly certain event. The study and the definition of this notion is therefore the subject of this subsection.

Searching for such a definition, we note that the following two properties seem to be appropriate for a nearly certain event: (1) it occurs nearly certainly in a single trial of the experiment, and (2) The event occurs nearly always in a long sequence of trials of the experiment. We note that property (1), unlike property (2), cannot be adopted as an operational definition of this kind of events because if a suitable event occurs in a single trial of the experiment, one cannot conclude that the event is nearly certain. Consequently, property (2) is adopted as the empirical definition of nearly certain events, while property (1) is considered by principle a property satisfied by these events. Let us explicitly formulate this conclusion in the following two statements:

\bg{defn} \label{neacer} An event is said to be \tit{nearly certain} if it occurs nearly always in a long sequence of trials of the experiment.
\en{defn}

\bg{princ}[Single trial] A nearly certain event occurs nearly certainly in a single trial of the experiment.
\en{princ}

The definition of a nearly certain event does not require that its long-run relative frequency is stable, provided it stays close to 1. Moreover, the vagueness of the definition of nearly certain event does not prevent us from formulating the following rigorous lemma:

\bg{lem} \label{nodis} Two disjoint events cannot both be nearly certain.
\end{lem}
\bg{proof} It is not possible for two disjoint events to occur nearly always in a long sequence of trials.
\end{proof}

\subsection{Probabilistic models and probabilistic experiments \label{promod}}

We can now define the conditions under which a probability space $(\Om, \AA, P)$ can be considered the probabilistic model of an experiment $\mE$.

For this purpose, let us first introduce the following simple notation: for $n \geq 1$, let the triple $(\Om^n, \AA^n, P^n)$ denote the $n$-fold product\footnote{where $\Om^n$ is the $n$-fold Cartesian product of $\Om$, $\AA^n=\si(\{A_1 \times \cdots \times A_n: A_i \in \AA \tx{ for }i=1, \ldots n\})$, and $P^n$ is the $n$-fold product measure of $P$, namely, the unique probability measure on $\AA^n$ such that $P^n(A_1 \times \cdots \times A_n)=P(A_1) \cdots P(A_n)$.} of the probability space $(\Om, \AA, P)$, and let $\mE^n$ denote the experiment composed of $n$ trials of the experiment $\mE$.

The first obvious requirement for a probability space $(\Om, \AA, P)$ to be a probabilistic model of a suitable experiment $\mE$ is that $\Om$ and $\AA$ can be identified, at least ideally, with the sample space and the class of the events of $\mE$, respectively. In this case, we say that the measurable space $(\Om, \AA)$ is a model of the event space of $\mE$. Moreover, it is assumed that if $(\Om, \AA)$ is a model of the event space of an experiment $\mE$, then $(\Om^n, \AA^n)$ is a model of the event space of $\mE^n$.

Let us then give the following definitions:

\bg{defn} (a) A probability space $(\Om, \AA, P)$ is a \tit{probabilistic model} of an experiment $\mathfrak{E}$ if $(\Om, \AA)$ is a model of the event space of $\mathfrak{E}$, and if the implication 
\bg{equation} \label{dpm}
P^n(A^{(n)}) \app 1 \Ra A^{(n)} \tx{ is nearly certain}
\en{equation}
holds for all $n \geq 1$, where $A^{(n)}$ is an element of $\AA^n$ and an event of $\mathfrak{E}^n$.

(b) An experiment is said to be a \tit{probabilistic experiment} if it admits a probabilistic model.
\end{defn}

\bg{prop} \label{prope1} An experiment admits at most one probabilistic model.
\en{prop}
\bg{proof} Let us recall Bernoulli's theorem in a form that is suitable for our purposes. Given a probability measure $P$ on $(\Om, \AA)$, a set $A \in \AA$, and $\ep >0$, let us define the following set of $\AA^n$:
\bg{equation}
S^\n(P,A,\ep):=\{\om^\n \in \Om^n: \vert P(A) - f_n (A, \om^\n) \vert < \ep\},
\end{equation}
where
\bg{equation}
f_n(A, \om^\n):= \frac{\sum_{i=1}^n\1_A(\om_i)}{n},
\end{equation}
and $\1_A$ is the characteristic function of the set $A$. In other words, the function $f_n(A, \om^\n)$ returns the relative frequency of the event $A$ in the sequence $\om^\n = (\om_1, \ldots, \om_n) \in \Om^n$, and the event $S^\n(P,A,\ep)$ is composed of the elements of $\Om^n$ for which the relative frequency of $A$ differs from $P(A)$ less than $\ep$.

Bernoulli's theorem states that (see for example \cite{papoulis1990probability}):
\bg{equation}
\lim_{n \to \ity} P^\n[S^\n(P,A,\ep)] \to 1,
\end{equation} 

Let us now prove the proposition. Let us suppose that an experiment admits two different probabilistic models; these two models necessarily have the same measurable space $(\Om, \AA)$ because both of these spaces are identified with the same event space. Let $P_1$ and $P_2$ be the probability measures of the two models, and let us suppose that there is $A \in \AA$ such that $P_1(A) \neq P_2(A)$. Let us define $\ep :=\vert P_1(A) - P_2(A)\vert/3$. The events $S^\n(P_1,A, \ep)$ and $S^\n(P_2,A, \ep)$ are disjoint for all $n$, and, from Bernoulli's theorem, $P_1^\n[S^\n(P_1,A, \ep)], P_2^\n[S^\n(P_2,A, \ep)] \app 1$ for $n$ large enough. If the probability spaces $(\Om, \AA, P_1)$ and $(\Om, \AA, P_2)$ were both probabilistic models of the same experiment, for $n$ large enough, the two disjoint events $S^\n(P_1,A, \ep)$ and $S^\n(P_2,A, \ep)$ would both be nearly certain, which is impossible due to lemma \ref{nodis}. This implies that $P_1=P_2$. 
\end{proof}

\bg{prop} \label{prope2} If $(\Om, \AA, P)$ is the probabilistic model of an experiment $\mathfrak{E}$, then (a) nearly certainly the relative frequency of any event $A$ of $\mE$ tends to $P(A)$ as the number of trials tends to infinity; (b) if an event $A$ of $\mE$ is nearly certain, then $P(A)$ is close to $1$.
\end{prop}
\bg{proof} (a) For all $A \in \AA$ and $\ep>0$, we have $P^\n[S^\n(P,A,\ep)] \app 1$ for $n$ sufficiently large, and therefore, the event $S^\n(P,A,\ep)$ is a nearly certain event of $\mE^n$.

(b) If $A$ is nearly certain, by definition, the long run relative frequency of $A$ is close to $1$, and from the previous point, one deduces that nearly certainly the relative frequency of $A$ is close to $P(A)$, so that $P(A)$ is close to $1$.
\end{proof}

Point (a) corresponds to the well-known association between probability and relative frequency; analogous proofs can be given for other properties of relative frequencies, such as the central limit theorem or the law of the iterated logarithm. Point (b) is generally not considered in the literature. Regarding this, let us remark that the definition of probabilistic model requires only the one-way implication $P(A) \app 1 \Ra $``$A$ is nearly certain'', while the opposite implication ``$A$ is nearly certain'' $\Ra P(A) \app 1$ is derived. Later, we show that the latter implication cannot be derived in the case of a generic typicalistic experiment.

\section{Typicalistic randomness \label{typran}}

As stated in the Introduction, a typicalistic experiment is an experiment that is possibly nonprobabilistic  but that nevertheless possesses a class of nearly certain events. In other words, in general, the long run relative frequency of the events of the experiment is nonstable, but for a suitable class of them it stays close to 1.

As a probabilistic experiment is modelled by a probability space, a typicalistic experiment can be modelled by a \tit{typicality space}, which is basically a probability space in which the probability measure is replaced by a less structured \tit{typicality measure} $T$. Typicality spaces and typicalistic experiments are related by the usual condition
\[
T(A) \app 1 \Ra A \tx{ is nearly certain}.
\]
Unlike the probabilistic case, in this case, this implication is not required to hold for the $n$-fold products of the typicality spaces and the experiments (see however remark \ref{remrip} below). All this is now presented in detail.

\subsection{Typicality spaces}

Let us first define a typicality space and then explain and justify its structure.

\bg{defn} \label{typmea} A \tit{typicality space} is the quadruple $(\Om, \AA, \EE, T)$, where $(\Om, \AA)$ is a measurable space, $\EE$ is a subset of $\AA$ containing at least the set $\Om$, and $T$ is the \tit{typicality measure}, that is, a set function $T:\EE \to (-\ity, 1]$ such that $T(\Om)=1$, satisfying the following \tit{consistency condition}:
\bg{equation} \label{concon}
\exists P \in \PP: T(A) \app 1 \Ra P(A) \app 1 \tx{ for all } A \in \EE,
\end{equation}
where $\PP$ is the class of the probability measures defined on $\AA$.
\end{defn}
Moreover, we say that the typicality space is \tit{nontrivial} if $T$ defines at least one typical set different from $\Om$, i.e., if there is at least one set $A \in \EE$ such that $T(A) \app 1$ and $A \neq \Om$. 

A consistent typicality measure has much less structure than a probability measure, namely, (i) it is not necessarily $\si$-additive; (ii) for this reason, it is not necessarily defined on a $\si$-algebra but rather on a generic class $\EE$; and (iii) it must be less than or equal to 1, but it may be negative since the only significant values of the measure are those close to 1. 

The consistency condition (\ref{concon}) can be easily motivated as follows. Let us consider a suitable experiment and let $f_n(A)$ denote the observed relative frequency of the event $A$ in a sequence of $n$ trials. The set function $f_n(\cdot):\AA \to [0,1]$ defines a (additive) probability measure on $\AA$. If $A$ is nearly certain and $n$ is sufficiently large, by definition, it must be $f_n(A) \app 1$. If the typicality space represents a suitable experiment (see later), we have the following chain of implications:
\[
T(A) \app 1 \Ra A \tx{ is nearly certain } \Ra f_n(A) \app 1,
\]
that is essentially condition (\ref{concon}). Condition (\ref{concon}) prevents, for example, that $T(A), T(B) \app 1$ for two disjoint sets $A$ and $B$, as expected.

To be precise, the conditions $\Om \in \EE$ and $T(\Om)=1$ are not justified by the previous reasoning, and they have been imposed mainly for aesthetic reasons. 

As one can see, the consistency condition (\ref{concon}) is vague. For this reason, rather than a formal mathematical condition, it should be considered as a criterion for verifying that a suitable set function derived from our theory can be consistently interpreted as a typicality measure.

\subsection{Typicalistic experiments}

Typicalistic experiments are defined analogously to probabilistic experiments, i.e., by first defining a typicalistic model and then defining as typicalistic an experiment that admits a typicalistic model.

\bg{defn} (a) A typicality space $(\Om, \AA, \EE, T)$ is a \tit{typicalistic model} of an experiment $\mathfrak{E}$ if $(\Om, \AA)$ is a model of the event space of $\mathfrak{E}$, and if 
\bg{equation} \label{dpmt}
T(A) \app 1 \Ra A \tx{ is nearly certain}.
\en{equation}

(b) An experiment is said to be a \tit{typicalistic experiment} if it admits a nontrivial typicalistic model.
\end{defn}

The requirement that the model of a typicalistic experiment must be nontrivial descends from the fact that a trivial typicality space, namely, a typicality space whose only typical set is the sample space $\Om$, is a typicalistic model of any experiment.

In the typicalistic case, propositions analogous to propositions \ref{prope1} and \ref{prope2} of the probabilistic case cannot be deduced. The consequence is that (1) an experiment may have many typicalistic models, (2) the inverse implication ``$A$ is nearly certain'' $\Ra T(A) \app 1$ may not hold in general, and (3) it is not possible to establish a precise correspondence between the value of $T(A)$ and the long run relative frequency of $A$. The only approximate correspondence between $T(A)$ and the long run relative frequency of $A$ occurs in the case in which $T(A) \app 1$ and derives from the definition of a nearly certain event; in this case, the long run relative frequency of $A$ is also close to 1.

\bg{rem} \label{remrip}
It is possible that further studies suggest extending condition (\ref{dpmt}) to the $n$-fold product of the typicality space and of the experiment, as in the probabilistic case.  In this case, the natural definition of the $n$-fold products of $\EE$ and $T$ is $\EE^n:=\{A_1 \times \cdots \times A_n: A_i \in \EE \tx{ for } i=1, \ldots, n\}$, and $T^n(A_1 \times \cdots \times A_n):=T(A_1) \cdots T(A_n)$, respectively. This definition, however, does not guarantee that $T^n$ is consistent if $T$ is consistent, and this requirement must be included in the definition of a consistent typicality measure. In other words, condition (\ref{concon}) must be replaced by the condition 
\bg{equation}
\exists P^\n \in \PP^\n: T^n(A_1 \times \cdots \times A_n) \app 1 \Ra P^\n(A_1 \times \cdots \times A_n) \app 1
\end{equation}
for all $n \geq 1$, where the adopted notation should be obvious.

We note, however, that propositions \ref{prope1} and \ref{prope2} of the probabilistic case cannot be deduced in the typicalistic case even if conditions (\ref{concon}) and (\ref{dpmt}) are extended to the $n$-fold products because the structure of the set $\EE^n$ on which $T^n$ is defined is too poor for deducing something such as Bernoulli's theorem. 
\end{rem}

\subsection{Examples}

\bg{example} Due to the vagueness of the consistency condition (\ref{concon}), the square or the square root of a probability measure can be considered typicality measures. More precisely, if $(\Om, \AA, P)$ is a probability space, then $(\Om, \AA, \AA, T)$, where $T=P^2$ or $T= \sqrt{P}$, are typicality spaces.

In fact, let us suppose that $T=\sqrt{P}$, and let $T(A) \app 1$, that is, $T(A) \geq 1- \ep$, with $\ep \ll 1$. Then, $P(A) \geq (1- \ep)^2 = 1 - 2 \ep + O(\ep^2) \app 1$. Analogously, if $T=P^2$ and $T(A) \geq 1- \ep$, then $P(A) \geq \sqrt{1- \ep} = 1 -  \ep/2 + O(\ep^2) \app 1$.
\end{example}

\bg{example} Let us consider a quantum particle in the state $\Psi$ at time $0$, evolving according to the time evolution operator $U(t)$. Let moreover $(\X, \XX)$ be its configuration space, where $\X:=\R^3$ and $\XX$ is the Borel $\si$-algebra of $\X$ and let $E$ be the associated projection valued measure (PVM). Let us define a typicality space as follows: $\Om:=\X \times \X$, $\AA:=\XX \times \XX$, $\EE:=\{(X \times Y: X, Y \in \XX\}$, and finally 
\[
T(X \times Y):=\|E(X)\Psi\|^2 \|E(Y) U(t)\Psi\|^2,
\]
where $t>0$ is fixed.

To prove that $T$ is consistent, we can utilize the probability measure $P$ on $(\X \times \X, \XX \times \XX)$ obtained by extending $T$ to $\XX \times \XX$  by $\si$-additivity. The consistency condition is trivially satisfied because $T(X \times Y) \app 1 \Ra P(X \times Y) = T(X \times Y) \app 1$.

This example has an obvious physical interpretation. If we assume that the particle has a position even in the absence of any measurement (see also Section \ref{quaevo}), the value  $\|E(X)\Psi\|^2$ (resp. $\|E(Y) U(t)\Psi\|^2$) can be considered the probability that the particle \tit{is} in region $X$ at time $0$ (resp. in region $Y$ at time $t$). The condition $T(X \times Y) \app 1$ implies $\|E(X)\Psi\|^2, \|E(Y) U(t)\Psi\|^2 \app 1$, and this allows us to deduce that nearly certainly the particle is in $X$ at the time $0$ \tit{ and } in $Y$ at the time $t$, even in absence of any measurement on the particle. This conclusion, although trivial, cannot be deduced from the orthodox interpretation of quantum mechanics.
\end{example}

\subsection{Mutual typicality}

In the Introduction, we proposed representing the evolution of a quantum particle by means of a typicality measure. A more appropriate representation requires what is referred to as a measure of \tit{mutual typicality}, which is a simple generalization of the former. In this section, the theory of mutual typicality is presented.

Let us recall first what mutual typicality means in probability theory. Given a probability space $(\Om, \AA, P)$, we say that two sets $A, B \in \AA$ are \tit{mutually typical} if 
\bg{equation} \label{mp}
M_P(A, B) := \frac{P(A \cap B)}{P(A \cup B)} \app 1,
\en{equation}
with the convention that $M_P(A, B)=1$ if $P(A \cup B) = 0$.

\bg{rem} Expression (\ref{mp}) makes sense even if $P$ is not a finite measure (for example, if $P$ is the Lebesgue measure on the real line), provided $P(A \cup B) < \ity$. An example of this kind is presented below.
\end{rem}

On the empirical side, we can generalize the definition of a nearly certain event by saying that two events $A$ and $B$ of an experiment are \tit{mutually nearly certain} if the event $A \cap B$ occurs in almost all the trials in which $A \cup B$ occurs. Analogous to the case of nearly certain events, we state the principle that if $A$ and $B$ are nearly mutually certain events, then nearly certainly $A$ occurs if and only if $B$ occurs is a single trial of the experiment.

One can prove that if $(\Om, \AA, P)$ is the probabilistic model of an experiment, then the following implication holds:
\bg{equation}
M_P(A,B) \app 1 \Ra A \tx{ and } B \tx{ are mutually nearly certain}.
\end{equation}

The notions of mutual typicality and mutual near certainty can be extended to nonprobabilistic randomness in the usual way:

\bg{defn} A \tit{mutual typicality space} is the quadruple $(\Om, \AA, \EE_2, M)$, where $(\Om, \AA)$ is a measurable space, $\EE_2$ is a nonempty subset of $\{(A,B): A, B \in \AA\}$, and $M$ is the \tit{mutual typicality measure}, that is, a set function $M:\EE_2 \to (-\ity, 1]$ satisfying the following \tit{consistency condition}:
\bg{equation} \label{mt}
\exists P \in \PP: M(A, B) \app 1 \Ra M_P(A, B) \app 1,
\end{equation}
where $\PP$ is the class of probability measures defined on $\AA$.
\end{defn}
In this case, we say that a mutual typicality space is \tit{nontrivial} if there is at least one pair $(A, B) \in \EE_2$, with $A \neq B$, such that $M(A,B) \app 1$.

The set function $M$ can fail to be consistent if, for example, it does not satisfies the following implication:
\bg{equation}
M(A_1, B_1),M(A_2, B_2) \app 1 \Ra M(A_1 \cup A_2, B_1 \cup B_2) \app 1.
\end{equation}
In fact this implication is satisfied by the set function $M_P$ for any probability measure $P$ because one can prove that
\bg{equation}
M_P(A_1, B_1) + M_P(A_2, B_2) \leq 1 + M(A_1 \cup A_2, B_1 \cup B_2).
\end{equation}

\bg{rem} \label{remN} For reasons that we present later, the condition
\bg{equation} \label{mt2}
\exists P \in \PP: N(A, B) \app 1 \Ra N_P(A, B):=\frac{2 P(A \cap B) }{P(A) + P(B)} \app 1
\end{equation}
instead of condition (\ref{mt}) is possibly considered as the consistency condition for the typicality measure $N$. Within the approximation allowed by the vagueness of the definition, the conditions (\ref{mt}) and (\ref{mt2}) are equivalent. To see this, we note first that 
\[
M_P=g(N_P),
\]
where $g:[0,1] \to [0,1]$ is the increasing bijection:
\bg{equation} \label{g}
g(\lam) = \frac{\lam}{2-\lam}, \tx{ with } g^\mo(\lam):=\frac{2 \lam}{1+\lam},
\end{equation}
where $g^\mo$ is also increasing. In fact, 
\bg{align*}
& \frac{1}{M_P(A, B) } = \frac{P(A \cup B) }{P(A \cap B)} =
 \frac{P(A) + P(B) - P(A \cap B) }{P(A \cap B)} = \\
& \frac{2}{N_P(A, B)} -1 = \frac{2 - N_P(A,B)}{ N_P(A,B)}.
\end{align*}
Moreover, $g(1-\ep) = 1-2 \ep + O(\ep^2)$  and $g^\mo(1-\ep) = 1- \ep/2 + O(\ep^2)$. Therefore, $\lam \geq 1- \ep$ implies that $g(\lam) \geq g(1- \ep) = 1- 2 \ep + O(\ep^2)$ and $g^\mo(\lam) \geq g^\mo(1-\ep) = 1 - \ep/2 + O(\ep^2)$, so that 
\[
M_P(A,B) \app 1 \Lra N_P(A,B) \app 1.
\]
In conclusion, the two conditions (\ref{mt}) and (\ref{mt2}) are equivalent.
\end{rem}

The relation between mutual typicality spaces and experiments is expressed by the following definition:

\bg{defn} (a) A mutual typicality space $(\Om, \AA, \EE_2, M)$ is a \tit{mutual typicalistic model} of an experiment $\mathfrak{E}$ if $(\Om, \AA)$ is a model of the event space of $\mathfrak{E}$, and if 
\bg{equation} \label{dpm2}
M(A, B) \app 1 \Ra A \tx{ and } B \tx{ are mutually nearly certain}.
\en{equation}

(b) An experiment is said to be a \tit{mutually typicalistic experiment} if it admits a nontrivial mutual typicalistic model.
\end{defn}

Analogous to the case of simple typicalistic randomness, an experiment may have many mutual typicalistic models, and it may happen that two events $(A, B) \in \EE_2$ are mutually nearly certain but $M(A, B) \not \app 1$.

\subsubsection{Example} \label{exa2}

Let us define the mutual typicality space of this example as follows: $\Om := \R \times \R$, $\AA:=\BB \times \BB$, where $\BB$ is the Borel $\si$-algebra of $\R$, and 
\bg{equation}
\EE_2:=\{ (X \times \R, \R \times Y): X, Y \tx{ are finite intervals of } \R \tx{ of the same length}\}.
\end{equation}
We note that $X \times \R \cap \R \times Y=X \times Y$. To define the mutual typicality measure, let us first introduce the following signed measure $Q$ on $(\R \times \R, \BB \times \BB)$: 
\bg{equation}
Q(A):= \int_A \Re K(x-y) dy \, dx,
\end{equation}
where 
\bg{equation} \label{defk} 
K(x) := \frac{1}{\sqrt{ 4 i \pi k t} } e^{i x^2/ 4 k t } \tx{ and } 
\Re \!K(x) = \frac{1}{\sqrt{8 \pi k t}} [\cos(x^2/ 4 k t) + \sin(x^2/ 4 k t)],
\end{equation}
and $t$ and $k$ are two positive parameters. We note that $K$ is basically the Feynman propagator of a free quantum particle, where $t$ represents the time. One can easily prove that $\int_\R K(x) dx = 1$, from which one deduces that $Q(X \times \R)=\ell(X)$, where $\ell$ is the Lebesgue measure. By means of $Q$, let us define the candidate mutual typicality measure 
\bg{equation} \label{defnq}
N_Q(X \times \R, \R \times Y) := \frac{2 Q(X \times \R \cap \R \times Y)}{Q(X \times \R) + Q(\R \times Y)}=\frac{2 Q(X \times Y)}{\ell(X) + \ell(Y)}.
\end{equation}
This expression has the same structure as expression (\ref{mt2}), with the difference that it is based on a signed measure rather than a normal measure. For this reason, it is necessary to prove that $N_Q \leq 1$. 

To prove this, note first that the operator 
\bg{equation}
[U \psi](x) :=\int_\R K(x-y) \psi(y) dy, \; \psi \in L^2(\R),
\end{equation}
is a unitary operator on $L^2(\R)$. Then, we can write
\bg{equation}
Q(X \times Y) = \Re \la \1_Y \vert U \1_X \ra \tx{ and } Q(X \times \R) = \|\1_X\|^2,
\end{equation}
where $\1_X$ and $\1_Y$ are the characteristic functions of the sets $X$ and $Y$, respectively (note that $\1_X, \1_Y \in L^2(\R)$ because $\ell(X), \ell(Y) < \ity$). Thus,
\bg{equation}
N_Q(X \times \R, \R \times Y) = 
\frac{2 \Re \la \1_Y \vert U \1_X \ra}{\|\1_Y\|^2 + \| \1_X\|^2}.
\end{equation}
One deduces that $N_Q \leq 1$ from the equality
\bg{equation}
\frac{2 \Re \la \1_Y \vert U \1_X \ra}{\|\1_Y\|^2 + \| \1_X\|^2} = 1 - \frac{\|\1_Y - U \1_X \|^2 }{\|\1_Y\|^2 + \| \1_X\|^2}.
\end{equation}
From this equality, one also deduces that the value of $N_Q(X \times \R, \R \times Y)$ measures the relative difference between the states $U \1_X$ and $\1_Y$, where the maximum value $1$ is achieved when $U \1_X=\1_Y$, and the value decreases as the relative difference increases.

The consistency of $N_Q$, namely, condition (\ref{mt2}), is proved (in a semiqualitative way) by means of the probabilistic expression $N_P$ based on the nonnegative measure
\bg{equation}
P(A):=\int_A \de(x-y) dydx.
\end{equation}
We note that $P(X \times Y) = \ell(X \cap Y)$. We have:
\[
N_P(X \times \R, \R \times Y) = \frac{2 P(X \times \R \cap \R \times Y)}{P(X \times \R)  + P(\R \times Y)}=\frac{2 \ell(X \cap Y)}{\ell(X) + \ell(Y)}.
\]
Let us explicitly calculate the values of $N_Q$ and $N_P$ for $X=[x_1, x_2]$, where $x_1 \leq x_2 \in \R$, and $Y=[x_1+a, x_2+a]$, where $a \in \R$. We have:
\bg{align*}
& Q(X \times Y) = \int_{X \times Y} \Re K(x-y) dx dy =\int_{x_1}^{x_2} \int_{x_1+a}^{x_2+a} \Re K(x-y) dy dx = \\
& = \Re [-K_2(a) + K_2(x_2-x_1 +a) + K_2(x_1-x_2+a) - K_2(a)],
\end{align*}
where $K_2$ is the second antiderivative of $K$. The real part of $K_2$ is:
\[
\Re K_2(x) = \frac{x}{2}[C(x/\sqrt{2 \pi k t}) + S(x/\sqrt{2 \pi k t})] + \sqrt{\frac{kt}{2\pi}} 
[\cos(x^2/(4 kt)) - \sin(x^2/(4 kt))], \\
\]
where $C$ and $S$ are the Fresnel integrals. Let us introduce the function
\bg{equation}
f(z) := \frac{z}{2}[C(z/\sqrt{2 \pi}) + S(z/\sqrt{2 \pi})] +  [\cos(z^2/4) - \sin(z^2/4)]/\sqrt{2 \pi},
\end{equation}
so that $\Re K_2(x) = \sqrt{k t} \cdot f(x/\sqrt{k t})$. Let us introduce the following variables:
\bg{equation}
d:= (x_2-x_1) /\sqrt{k t} \tx{ and } h:= a /\sqrt{k t}.
\end{equation}
We then have:
\bg{equation}
N_Q(X \times \R, \R \times Y) = \frac{f(h+d) + f(h-d) -2 f(h)}{d}.
\end{equation}

The expression of $N_P$ in terms of the parameters $d$ and $h$ is much more easily calculated, and it is 
\bg{equation}
N_P(X \times \R, \R \times Y) = \frac{\max\{0, d-  \vert h \vert\}}{d}.
\end{equation}

\begin{figure}
\centering
\includegraphics{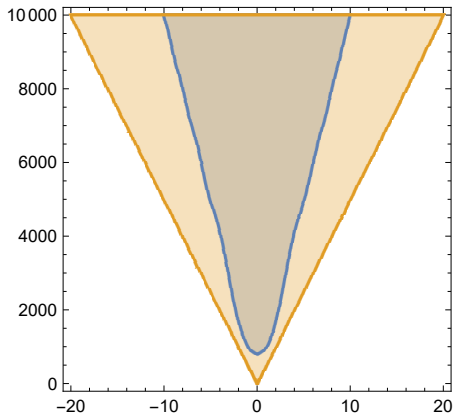}
\caption{This graphic has been obtained by the command RegionPlot of Wolfram Mathematia. In the blue region $N_Q \geq 0.999$, and in the yellow region $N_P \geq 0.998$. The values of $N_Q$ and $N_P$ are functions of the variables $d$ (ordinate) and $h$ (abscissa). The form of the two regions supports the conjecture that $N_Q(h, d) \app 1 \Ra N_P(h,d) \app 1$ also beyond the values of $h$ and $d$ shown in the figure.}
\label{figcomp}
\end{figure}

The condition of consistency 
\bg{equation}
N_Q(X \times \R, \R \times Y) \app 1 \Ra N_P(X \times \R, \R \times Y) \app 1.
\end{equation}
is supported graphically by Fig. \ref{figcomp}. This figure shows that the region in which $N_Q \geq 0.999$ is contained in the region in which $N_P \geq 0.998$.

This example is rather elementary, mainly due to the constraints that the sets $X$ and $Y$ have the same length and that they are intervals. These constraints have been imposed to allow us to provide a demonstration, albeit semiqualitative, of the consistency of $N_Q$. It seems reasonable to conjecture that $N_Q$ remains consistent even if the first constraint is removed. On the other hand, if the constraint that $X$ and $Y$ are intervals is removed, arguably $N_Q$ is no longer consistent due to the possible presence of interference phenomena. The quantum typicality measure presented in Section \ref{typspa}, although based on a similar quantum formalism, has a more restrictive structure than that presented in this example (see also remark \ref{rempre}).

\section{Quantum evolution as a typicalistic phenomenon \label{quaevo}}

This section begins the second part of the paper in which the hypothesis is developed that the evolution of a system of quantum particles is a typicalistic phenomenon. This possibility can be considered a natural solution to the problem that the quantum formalism does not allow us to derive a probability measure on a set of trajectories, as explained in this introductory section. For simplicity, the formalism is developed for the case of a single particle, but it can be trivially extended to the case of a system of particles.

Let $\HH:=L^2(\R^3)$ be the Hilbert space of a nonrelativistic particle, $\Psi$ its normalized state at the initial time $t=0$, $U(t,s)$ the unitary time evolution operator from the time $s$ to the time $t$, $E$ the PVM associated with the position operators, and $E_t:=U^\dg(t,0)EU(t,0)$ the PVM associated with the position operators at the time $t$ (in Heisenberg representation). Both $E$ and $E_t$ are defined on the measurable space $(\X, \XX)$, where $\X=\R^3$ is the configuration space of the particle and $\XX$ its Borel $\si$-algebra. The evolution of the particle is considered during a suitable time interval $T:=[0,t_F)$, during which no measurement is performed on the particle.

It is well-known that for any physically reasonable Hamiltonian, the PVMs $E_t$ and $E_s$ do not commute for $t \neq s$. For a moment, let us assume instead that they commute, and let us study the consequences. For $X_1, \ldots, X_n \in \XX$ and $t_1 < \cdots < t_n \in T$, the expression
\bg{equation} \label{fdd}
P_{t_1, \ldots, t_n} (X_1, \ldots, X_n):=\la \psi \vert E_{t_1}(X_1)  \cdots E_{t_n}(X_n)\vert\psi\ra
\en{equation}
would be positive definite and $\si$-additive for any of the arguments $E_{t_i}(\cdot)$ and could be interpreted as the probability that the particle is in region $X_1$ at time $t_1$ and in region $X_2$ at time $t_2$ and $\cdots$ in region $X_n$ at time $t_n$. The set of the expressions of this type would form a system of finite dimensional distributions that, according to Kolmogorov extension theorem, could be extended to a probability measure $P$ on the set $\X^T$ of all the possible trajectories of the particle during the interval $T$. Consequently, one could assume that the particle follows a trajectory chosen at random from $\X^T$ according to the probability $P$, analogously, for example, to a particle subjected to Brownian motion. By applying this approach to the whole universe (considered a system of quantum particles), the measurement problem and the problem of the quantum-to-classical transition would disappear.

Unfortunately, the projections $\{E_t\}_{t \in T}$ do not commute for different times, the expression (\ref{fdd}) is not a finite dimensional distribution, and the quantum formalism does not allow us to endow the set of trajectories of the particle with a probability measure. This may be one of the reasons that led some of the founders of quantum mechanics to deny that the quantum particles follow definite trajectories, as expressed by the well known claim: ``the particle does not have a definite position until it is measured''. However, the fact that a theory does not allow us to define a probability measure on a suitable set does not imply that the set does not exist. The proposal is then to assume that the particle keeps following a definite trajectory (i.e., the set of trajectories still exists) but that its evolution exhibits the kind of randomness that must be represented by a typicality measure rather than by a probability measure, as explained in the Introduction. In the next section, a physical principle from which to derive this typicality measure (actually, a mutual typicality measure) is proposed:  We refer to as \tit{the principle of separation}.

\subsection{Separators and the principle of separation \label{prinsep}}

This principle is based on the example of the beam splitter presented in the Introduction, which is in turn an approximate version of the so-called \tit{Einstein's Boxes} thought experiment \cite{norsen2005einstein}. Let us briefly recall this experiment.

At time $t_1$ a particle is prepared in such a way that its wave function is subdivided into two separated impenetrable boxes. The two boxes are then transported to two very distant places. At a subsequent time $t_2$, the boxes are opened to determine in which of these the particle is contained. A natural assumption is that the particle was since the time $t_1$ in the same box in which it has been found at the time $t_2$, even if such a conclusion cannot be derived from standard quantum theory.

Let us formulate this experiment and the related assumption in a more precise way. Let $\K:=\X \times T$ and let $\KK$ be the Borel $\si$-algebra of $\K$. The elements of $\KK$ are referred to as \tit{tubes}, and for $K \in \KK$ and $t \in T$, the symbol $K_t$ denotes the section of the tube $K$ at  time $t$, i.e., $K_t:=\{x \in \X: (x, t) \in K\}$. Now, with reference to the Einstein's boxes thought experiment presented above, let us choose $K$ such that, for $t \in [t_1, t_2]$, the section $K_t$ corresponds to the region inside one of the boxes. One easily recognizes that the following equality holds:
\bg{equation} \label{ebox}
E(K_t)U(t,0) \Psi= U(t,s)E(K_s)U(s,0) \Psi
\en{equation}
for all $s,t \in [t_1, t_2]$. In the above equation, the action of the walls of the box on the wave function is incorporated in the time-dependent evolution operator $U(t,s)$. Equality (\ref{ebox}) derives from the fact that since the box is impenetrable, the part of the wave function inside the box at a suitable time $s \in [t_1, t_2]$ remains in the box under time evolution. This equality can be rewritten as
\bg{equation} \label{esep}
E_t(K_t) \Psi= E_s(K_s) \Psi \tx{ for all } s,t \in [t_1, t_2].
\en{equation}
The assumption mentioned above can be expressed by the following principle, whose name is motivated below:
\bg{pes} If condition (\ref{esep}) holds, then the particle is in region $K_{t_1}$ at time $t_1$ if and only if it is in region $K_{t_2}$ at time $t_2$.
\end{pes}

Due to the spreading of the wave function under Schr\"odinger evolution, the exact condition (\ref{esep}) is possible only in the presence of infinite potential wells, i.e., of an impenetrable box. In more realistic situations this condition can hold only approximately. A tentative approximate version of this equation is the following: 
\bg{equation} \label{uasep}
\|E_t(K_t) \Psi- E_s(K_s) \Psi\|^2 \app 0 \tx{ for all } s,t \in [t_1, t_2],
\en{equation}
where $x \app 0$ means $x \leq \ep$ with $\ep \ll 1$. This condition, however, is ambiguous, because it can be satisfied either because $E_t(K_t) \Psi \app E_s(K_s) \Psi$ or because $\|E_t(K_t) \Psi\|^2, \|E_s(K_s) \Psi\|^2 \app 0$. To remove this ambiguity the condition must be normalized, and it turns out that $\|E_t(K_t) \Psi\|^2+ \|E_s(K_s) \Psi\|^2$ is an appropriate normalization factor. The proposed approximate version of equation (\ref{esep}) is therefore 
\bg{equation} \label{nasep}
\frac{\|E_t(K_t) \Psi- E_s(K_s) \Psi\|^2}{\|E_t(K_t) \Psi\|^2+\|E_s(K_s) \Psi\|^2} \app 0 \tx{ for all } s,t \in [t_1, t_2],
\en{equation}
which is equivalent to the condition
\bg{equation} \label{nasep2}
\frac{2 \Re \la E_t(K_t) \Psi \vert E_s(K_s) \Psi \ra}{\|E_t(K_t) \Psi\|^2+\|E_s(K_s) \Psi\|^2} \app 1 \tx{ for all } s,t \in [t_1, t_2],
\en{equation}
where, we recall that $x \app 1$ means $x \geq 1 - \ep$ with $\ep \ll 1$. The elements of $\KK$ satisfying the above condition are so important in the approach proposed here that they are worth a specific definition:
\bg{defn} If a tube $K \in \KK$ satisfies condition (\ref{nasep2}) we say that it \tit{separates} (or is a \tit{separator} of) $\Psi$ in the time interval $[t_1, t_2]$.
\end{defn}
The existence of a separator reveals a meaningful pattern of the wave function, namely, the existence of a part of the wave function that is spatially separated from the rest of the wave function and that evolves independently from it; this is basically the reason for the name ``separator''. The typical example is the case in which the wave function splits into two (or more) nonoverlapping wave packets, as in the Einstein's Boxes thought experiment or in the experiment of the beam splitter presented in the Introduction. In this case, the (possibly approximate) spatial supports of any wave packet during the time interval in which it does not overlap the rest of the wave function are the sections of a separator of $\Psi$ in that time interval. However, the existence of a separator is not necessarily associated with the splitting of the wave function into nonoverlapping wave packets, as shown in example  \ref{exsep} below. This example also shows that the existence of separators seems to be a structural property of Schr\"odinger evolution, at least in the asymptotic regime.

As the principle of exact separation is associated with the exact condition (\ref{esep}), it is reasonable to associate an approximate version of such a principle with the approximate condition (\ref{nasep2}):

\bg{ps} If condition (\ref{nasep2}) holds, then nearly certainly the particle is in region $K_{t_1}$ at time $t_1$ if and only if it is in region $K_{t_2}$ at time $t_2$.
\end{ps}

In other words, this principle states that condition (\ref{nasep2}) implies that the two events: The particle is in region $K_{t_i}$ at time $t_i$, where $i=1,2$, are nearly mutually certain. A mutual typicality space implementing this principle is presented in the next section.

\bg{example} \label{exsep} In this example, it is shown that the existence of separators is a structural property of the asymptotic evolution of a quantum particle and that they are not necessarily related to the splitting of the wave function into nonoverlapping wavepackets.

Let us consider the evolution of a free particle of mass $m$ in the time interval $T=[0, \ity)$; the Hamiltonian is $H=-\De /2m$. Let $P$ be a measurable subset of $\R^3$, to be interpreted as a subset of the momentum space, and let
\bg{equation}
K:=\{(pt/m, t) \in \X \times [0, \ity): t \in [0, \ity) \tx{ and } p \in P\}.
\en{equation}
In words, $K$ is the union of the classical trajectories for $t \geq 0$ of free particles with momentum $p \in P$ which is in the origin at time $t=0$. Obviously $K_t=\{pt/m:p \in P\}=:Pt /m$. Let $F$ be the PVM associated with the momentum operators of the particle. We need the following
\bg{prop} 
\bg{equation} \label{lim1}
s-\lim_{t \to \ity} E_t(Pt/m) = F(P).
\en{equation}
\end{prop}
\bg{proof} Let us introduce Dollard's operators \cite{dollard1969scattering}:
\bg{align*}
& (C_t \psi)(x) := \sqrt{\frac{m}{i t}} e^{imx^2/2t} \tpsi(mx/t), \\
& (Q_t \psi)(x) :=  e^{imx^2/2t} \psi(x),
\end{align*}
where $\tpsi$ is the Fourier transform of $\psi$. They are unitary and satisfy the equation $e^{-iH t} =C_t Q_t$. Let us prove first that $C_t^\mo E(t P /m) C_t = F(P)$:
\bg{align*}
& [C_t F(P) \psi](x) = \sqrt{\frac{m}{it}} e^{imx^2/2t} \1_P(mx/t) \tpsi(mx/t) = \\
& \1_{t P/m}(x) \sqrt{\frac{m}{it}} e^{imx^2/2t}  \tpsi(mx/t) = [E(tP/m)C_t \psi](x),
\end{align*}
where $\1_A$ is the characteristic function of the set $A$. Thus,
\bg{align*}
& E_t(tP/m) = e^{iH t}E(t P/m) e^{-iHt} = \\
& Q^\mo_t C^\mo_t E(t P/m) C_t Q_t = Q^\mo_t F(P) Q_t,
\end{align*}
and the thesis descends from the fact that $s-\lim_{t \to \ity} Q_t = I$.
\end{proof}
From limit (\ref{lim1}) and the Cauchy convergence criterion, it follows that $\sup_{s,t \geq t_0}\|E(K_t)\Psi- E(K_s)\Psi \| \to 0$ for $t_0 \to \ity$, from which one easily deduces that there is $t_0 \in [0, \ity)$ such that condition (\ref{nasep}) (and therefore condition (\ref{nasep2}) as well) is satisfied for any $t, s \in [t_0, \ity)$.
\end{example}

\section{The quantum typicality measure \label{typspa}}

Let us now define the various elements of the mutual typicality space $(\Om, \AA, \EE_2, M)$ representing the evolution of a quantum particle; this space and the associated measure are referred to as the \tit{quantum typicality space} and the \tit{quantum typicality measure}, respectively. There is a certain stretch in this terminology because the fact that the quantum typicality measure satisfies the consistency conditions (\ref{mt}) or (\ref{mt2}) is not proven but only conjectured.

The sample space $\Om$ is $\X^T$, which consists of all the possible trajectories $\om:T \to \X$. Let $\pi_t: \Om \ni \om \to \om(t) \in \X$ be the canonical projection at time $t$. For $X \in \XX$, the set $\pi^\mo_t(X) \se \Om$ is composed of all the trajectories $\om$ such that $\om(t) \in X$, and therefore, it can be associated with the event ``the particle is in region $X$ at time $t$''. The sets/events of this kind are referred to as the \tit{elementary sets/events}, and the following more compact notation is adopted for denoting them:
\bg{equation}
\la t, X\ra:=\pi^\mo(X).
\en{equation}
Moreover, let  $\SS$ denote the class of the elementary sets, that is:
\bg{equation}
\SS:= \{\la t, X \ra: t \in T \tx{ and } X \in \XX\}.
\en{equation}
By means of the class $\SS$, we define $\AA$ and $\EE_2$ as follows:
\bg{equation}
\AA:= \si (\SS) \tx{ and } \EE:=\{(S_1, S_2): S_1, S_2 \in \SS\}.
\en{equation}
The \sa $\AA$ is the usual \sa generated by the cylinder sets (we recall that a cylinder set is a subset of $\Om$ of the type $\la t_1, X_1 \ra \cap \cdots \cap \la t_n, X_n \ra$).

Let us now define the mutual typicality measure $N$, which is supposed to satisfy the consistency condition (\ref{mt2}). The ``official'' mutual typicality measure $M$ is subsequently obtained from $N$ by means of the bijection (\ref{g}). The reason for this double passage is explained below.

For convenience, let us first introduce the following function:
\bg{equation} \label{mequ}
n(t, X, s, Y) := \frac{2 \Re \la E_s(Y)\Psi\vert E_t(X) \Psi \ra}{\|E_t(X)\Psi\|^2 + \|E_s(Y)\Psi\|^2},
\en{equation}
where $s,t \in T$, $X,Y \in \XX$, and the following convention is adopted:
\bg{equation} \label{conv}
n(t, X, s, Y):=0 \tx{ if } \|E_t(X)\Psi\|=\|E_s(Y)\Psi\|=0.
\en{equation}
We note that $n \leq 1$. Two further definitions are needed. First, let $I[t_1, t_2]$ denote the closed interval defined by $t_1$ and $t_2$, namely, $I[t_1,t_2]=[t_1,t_2]$ if $t_1 \leq t_2$, and $I[t_1,t_2]=[t_2,t_1]$ if $t_2 \leq t_1$. Second, for $t_1 \neq t_2 \in T$ and $X_1, X_2 \in \XX$, let us define
\bg{equation}
\KK[t_1, X_1, t_2, X_2]:= \{ K \in \KK: K_{t_1} = X_1 \tx{ and } K_{t_2} = X_2 \},
\en{equation}
where we recall that $\KK$ is the set of the tubes defined in section \ref{prinsep}. The mutual typicality measure $N$ can then be defined as follows:
\bg{equation} \label{mtmis}
N(\la t_1, X_1\ra,\la t_2, X_2\ra):= \left \la \bg{array}{ll}
\sup_{K \in \KK[t_1, X_1, t_2, X_2]} \inf_{t, s \in I[t_1, t_2]} n(t, K_t, s, K_s) & \tx{ for } t_1 \neq t_2, \\ \\
n(t_1, X_1, t_2, X_2) & \tx{ for } t_1 = t_2.
\en{array} \right. 
\en{equation}
Before explaining this definition, let us prove some simple properties for $N$:
\bg{prop} \hspace{1mm}
\bg{itemize}
\item[(a)] $N(\la t_1, X_1\ra,\la t_2, X_2\ra) \leq n(t_1, X_1, t_2, X_2)$;
\item[(b)] $\lim_{t_2 \to t_1} N(\la t_1, X_1 \ra,\la t_2, X_2\ra) = n(t_1, X_1, t_1, X_2)$;
\item[(c)] $N(\la t_1, X_1 \ra,\la t_2, \X \ra)= n(t_1, X_1, t_2, \X)= \frac{2 \| E_{t_1}(X_1)\|^2}{1 + \| E_{t_1}(X_1)\|^2}$;
\item[(d)] $N(\la t_1, X_1 \ra,\la t_2, \es \ra)= n(t_1, X_1, t_2, \es)=0$.
\en{itemize}
\en{prop}

\bg{proof} (a) If $t_1=t_2$ the thesis is true by definition. Let us suppose $t_1 \neq t_2$. In this case, 
\[
\inf_{t, s \in I[t_1, t_2]} n(t, K_t, s, K_s) \leq n(t_1, X_1, t_2, X_2)
\]
for all $K \in \KK[t_1, X_1, t_2, X_2]$, from which the thesis.

(b) Let $K \in \KK[t_1, X_1, t_2, X_2]$ be the tube such that $K_{t_1} = X_1$ and $K_t=X_2$ for $t \neq t_1$, and let us define 
\[
A_{t_1, t_2} := \inf_{s,t \in I[t_1, t_2]} n(s, K_s, t, K_t).
\]
We want to prove that
\bg{equation} \label{mlim}
\lim_{t_2 \to t_1} A_{t_1, t_2} = n(t_1, X_1, t_2, X_2).
\en{equation}
Assume first $t_2 > t_1$. We have
\bg{align*}
& A_{t_1, t_2} = \inf_{s,t \in [t_1, t_2]} n(s, K_s, t, K_t) = \\
& \min \left \{ \inf_{\substack{s=t_1 \\ t =t_1}} n(),  \inf_{\substack{s=t_1 \\ t \in (t_1, t_2]}} n(), \inf_{s,t \in (t_1, t_2]} n() \right \} =\\
& \min \left \{n(t_1, X_1, t_1, X_1),  \inf_{t \in (t_1, t_2]} n(t_1, X_1, t, X_2), \right.\\
& \left. \inf_{s,t \in (t_1, t_2]} n(s, X_2, t, X_2) \right \},
\end{align*}
where in the first passage the equality 
\[
\inf_{s, t \in I[t_1, t_2]} n(t, K_t, s, K_s)=\inf_{s < t \in I[t_1, t_2]} n(t, K_t, s, K_s),
\]
deriving in turn from the equality $n(t, K_t, s, K_s)=n(s, K_s, t, K_t)$, has been used. Since the limit of the minimum is the minimum of the limits, let us calculate the limits for $t_2 \to t_1^+$ of the three elements in the brackets. We have:
\bg{align*}
& \lim_{t_2 \to t_1^+} n(t_1, X_1, t_1, X_1) = n(t_1, X_1, t_1, X_1), \\
& \lim_{t_2 \to t_1^+} \inf_{t \in (t_1, t_2]} n(t_1, X_1, t, X_2) = n(t_1, X_1, t_1, X_2), \\
& \lim_{t_2 \to t_1^+} \inf_{s,t \in (t_1, t_2]} n(t, X_2, s, X_2)= n(t_1, X_2, t_1, X_2).
\end{align*}
The first limit is obvious. The second limit stems from the fact that 
\[
\lim_{t_2 \to t_1^+} \inf_{t \in (t_1, t_2]} n(t_1, X_1, t, X_2)=\lim_{t_2 \to t_1^+}  n(t_1, X_1, t, X_2),
\]
and in the case $\|E_{t_1}(X_1) \Psi\|=\|E_{t_1}(X_2) \Psi\|=0$, it is correct due to the convention (\ref{conv}). The third limit is obvious in the case $\|E_{t_1}(X_2) \Psi\| \neq 0$. On the other hand, in the case $\|E_{t_1}(X_2) \Psi\| = 0$, we have that $\inf_{s,t \in (t_1, t_2]} n(t, X_2, s, X_2)=0$, and the limit is still correct due to the convention (\ref{conv}).

One can easily verify that 
\[
\min \left \{n(t_1, X_1, t_1, X_1), n(t_1, X_1, t_1, X_2), n(t_1, X_2, t_1, X_2)\right \} = n(t_1, X_1, t_1, X_2).
\]
In fact, this equation is correct in the case in which $\|E_{t_1}(X_1) \Psi\|, \|E_{t_1}(X_2) \Psi\| > 0$ because in this case $n(t_1, X_1, t_1, X_1)=n(t_1, X_2, t_1, X_2)=1$, and it is correct in the case in which either $\|E_{t_1}(X_1) \Psi\|$ or $\|E_{t_1}(X_2) \Psi\|$ (or both) is or are equal to 0 because in this case $n(t_1, X_1, t_1, X_2)=0$.

We have therefore proved that $\lim_{t_2 \to t_1^+} A_{t_1, t_2} =n(t_1, X_1, t_1, X_2)$. With analogous reasoning, one can prove that $\lim_{t_2 \to t_1^-} A_{t_1, t_2}=n(t_1, X_1, t_1, X_2)$, and therefore, equation (\ref{mlim}) is proved. Moreover, one can easily prove $\lim_{t_2 \to t_1} n(t_1, X_1, t_2, X_2) = n(t_1, X_1, t_1, X_2)$.

In conclusion, from the inequalities
\[
A_{t_1, t_2} \leq N(\la t_1, X_1 \ra, \la t_2, X_2\ra) \leq n(t_1, X_1, t_2, X_2)
\]
and the squeeze theorem one obtains
\[
\lim_{t_2 \to t_1} N(\la t_1, X_1 \ra, \la t_2, X_2\ra) = n(t_1, X_1, t_1, X_2).
\]

(c) If $t_1=t_2$ the thesis is true by definition. Let $K \in \KK[t_1, X_1, t_2, \X ]$ be such that $K_{t_1} = X_1$ and $K_t=\X$ for $t \neq t_1$. One easily verify that 
\[
\inf_{s, t \in I[t_1, t_2]} n(s, K_s, t, K_t)= n(t_1, X_1, t_2, \X),
\]
which is correct also in the case in which $\|E_{t_1}(X_1)\Psi)\|=0$. So $N(\la t_1, X_1 \ra,\la t_2, \X \ra) \geq n(t_1, X_1, t_2, \X)$, and the thesis is proved by applying point (a).

(d) The proof is analogous to the previous case. In this case, $K_{t_1}=X_1$ and $K_t=\es$ for $t \neq t_1$, which implies $\inf_{s, t \in I[t_1, t_2]} n(s, K_s, t, K_t) = 0= n(t_1, X_1, t_2, \es)$.
\en{proof}

Point (b) motivates the definition of $N$ for the case $t_1 = t_2$. Moreover $\la t, \X \ra=\X^T$ and $\la t, \es \ra=\es$ for all $t \in T$, and therefore, the definition of $N$ is consistent only if $N(\la t_1, X_1 \ra, \la t_2, \X\ra)$ and $N(\la t_1, X_1 \ra, \la t_2, \es \ra)$ do not depend on $t_2$; this has been proven by points (c) and (d).

The motivation for the given definition of $N$ is simple: One easyly sees that the sets $\la t_1, X_1 \ra$ and $\la t_2, X_2 \ra$ are mutually typical according to $N$, that is, $N(\la t_1, X_1 \ra,\la t_2, X_2 \ra) \app 1$, if and only if there is a tube $K \in \KK[t_1, X_1, t_2, X_2]$ that separates $\Psi$ in the time interval $I[t_1, t_2]$. In this case, according to the principle of separation, the events $\la t_1, X_1 \ra$ and $\la t_2, X_2 \ra$ are nearly mutually certain, as expected. In other words, if $N$ is defined by equation (\ref{mtmis}), the principle of separation expresses the usual implication that if $\la t_1, X_1 \ra$ and $\la t_2, X_2 \ra$ are mutually typical, then the corresponding events are nearly mutually certain.

\bg{rem} \label{rempre} In our previous papers \cite{galvan:1, galvan:4}, the mutual typicality measure for a quantum particle was defined to be $n$ (equation (\ref{mequ})) rather than $N$ (equation (\ref{mtmis})). This definition has two drawbacks: (1) It does not truly implement the principle of separation because it does not require that the wave function separates during the whole time interval $[t_1, t_2]$; (2) in general, $n$ is not consistent, as can be shown by means of a suitable interference experiment of the type presented in \cite{reck1994experimental}. We do not go into the details here.
\end{rem}

\subsection{The quantum typicality measure and the Born probabilities \label{born}}

The quantum formalism introduced in Section \ref{quaevo} naturally defines the class of probability measures $\{B_t\}_{t \in T}$ on $(\X, \XX)$ as follows:
\bg{equation} \label{bornp} 
B_t(X):=\|E_t(X)\Psi\|^2.
\end{equation}
The probability measures $\{B_t\}_{t \in T}$ are referred to as \tit{the Born probabilities}. In our context, $B_t(X)$ can be interpreted as the probability that particle \tit{is} in region $X$ at time $t$. This interpretation differs from the usual Born rule, according to which $B_t(X)$ is the probability \tit{to find} the particle in region $X$ at time $t$ if a measurement is performed.

For equal times elementary sets, the typicality measure $N$ reads:
\[
N(\la t, X_1\ra, \la t, X_2\ra) = \frac{2 B_t(X_1 \cap X_2)}{B_t(X_1)+ B_t(X_1)}=N_{B_t}(X_1, X_2),
\]
and in particular
\[
N(\la t, X\ra, \la t, \X\ra) = \frac{2 B_t(X)}{1 + B_t(X)}.
\]
 
This is not completely satisfactory because we want the term on the right of the above equality to be $B_t(X)$. For this reason, the set function 
\bg{equation} \label{qtm}
M:= g(N)
\en{equation}
is adopted instead of $N$ as the ``official'' quantum typicality measure, where $g$ is the bijection defined by equation (\ref{g}). According to the reasoning of remark \ref{remN}, as mutual typicality measures, the set functions $M$ and $N$ are equivalent, that is, $M(\la t_1, X_1\ra, \la t_2, X_2 \ra) \app 1 \Lra N(\la t_1, X_1\ra, \la t_2, X_2 \ra) \app 1$, but one easily sees that $M$ has the required property that
\bg{equation}
M(\la t, X\ra, \la t, \X\ra)= B_t(X).
\end{equation}
In other words, for equal times sets, the quantum typicality measure $M$ incorporates the Born probabilities $\{B_t\}_{t \in T}$ and can be interpreted consequently.

\subsection{On the consistency of the quantum typicality measure}

The consistency condition for the quantum typicality measure $N$ reads:
\bg{equation} \label{coqtm}
\exists P \in \PP: N(\la t_1, X_1\ra, \la t_2, X_2\ra) \app 1 \Ra N_P(\la t_1, X_1\ra, \la t_2, X_2\ra) \app 1,
\end{equation}
where here $\PP$ is the class of the probability measures on $(\X^T, \si(\SS))$. The proof of condition (\ref{coqtm}) seems to be rather difficult, and it is not given in this paper. However, some possible simplifications of this problem are presented in this section.

A first simplification consists of replacing the global quantum typicality space defined in the previous sections with the class of typicality spaces $\{(\Om, \AA, \EE_2, N_{t_1, t_2})\}_{t_1, t_2 \in T}$ defined as follows: $\Om := \X \times \X$, $\AA:=\AA \times \AA$, $\EE_2:=\{(X \times \X, \X \times Y):X, Y \in \XX\}$, and 
\bg{equation}
N_{t_1,t_2}(X_1 \times \X, \X \times X_2):= N(\la t_1, X_1 \ra, \la t_2, X_2 \ra), 
\en{equation}
To prove the consistency of $N$, it is then sufficient to prove the consistency of $N_{t_1,t_2}$ for all $t_1, t_2 \in T$; more specifically, it is sufficient to prove that, for all $t_1, t_2 \in T$, 
\bg{equation} \label{ccr2}
\exists P_{t_1, t_2} \in \PP:  N_{t_1, t_2} (X_1\times \X, \X \times X_2) \app 1 \Ra N_{P_{t_1,t_2}}(X_1 \times \X, \X \times X_2) \app 1,
\end{equation}
where, in this case, $\PP$ is the class of the probability measures on $(\X \times \X, \XX \times \XX)$. This conclusion derives from the fact that a class of bidimensional distributions $\{P_{t_1, t_2}\}_{t_1, t_2 \in T}$ satisfying the above condition can be certainly extended to a probability measure $P$ on $(\X^T, \si(\SS))$) and that this extension necessarily satisfies condition (\ref{coqtm}). 

The second simplification is the following. From the function $n(t, X, s, Y)$ defined by equation (\ref{mequ}) let us derive the following function:
\bg{equation}
n_1(t_1, X_1, t_2) := \bg{cases}
\inf_{t \in I[t_1, t_2)} \sup_{X \in \XX} n(t_1, X_1, t, Y) & \tx{ for } t_1 \neq t_2,\\
1 & \tx{ for } t_1 = t_2.
\end{cases}
\end{equation}
We note that  $n_1(t_1, X_1, t_2) \app 1$ implies that, for all $t \in I[t_1, t_2)$, there is a set $X$ (depending on $t$) such that $n(t_1, X_1, t, X) \app 1$. One can prove that
\bg{prop} 
\bg{equation}
N_{t_1, t_2} (X_1 \times \X, \X \times X_2) \leq  \min \left \{n_2(t_1, X_1, t_2), n(t_1, X_1, t_2, X_2) \right \}.
\end{equation}
\end{prop}
\bg{proof} If $t_1 = t_2$, then $N_{t_1, t_2} (X_1 \times \X, \X \times X_2)= n(t_1, X_1, t_2, X_2)$, and the proof is obvious. Assume $t_1 < t_2$. Then:
\bg{align*}
& N_{t_1, t_2} (X_1 \times \X, \X \times X_2) = N(\la t_1, X_1 \ra, \la t_2, X_2 \ra) = \\
& = \sup_{K \in \KK[t_1, X_1, t_2, X_2]} \inf_{t, s  \in [t_1, t_2]} n(t, K_t, s, K_s) \leq \\
& \leq \sup_{K \in \KK[t_1, X_1, t_2, X_2]} \inf_{t \in [t_1, t_2]} n(t_1, X_1, t, K_t)  = \\
& = \min \left \{ \sup_{K \in \KK[t_1, X_1, t_2, X_2]} \inf_{t \in [t_1, t_2)} n(t_1, X_1, t, K_s), n(t_1, X_1, t_2, X_2) \right \}= \\
& =  \min \left \{\inf_{t \in [t_1, t_2)} \sup_{X \in \XX} n(t_1, X_1, t, X), n(t_1, X_1, t_2, X_2) \right \}= \\
& = \min \left \{n_1(t_1, X_1, t_2), n(t_1, X_1, t_2,X_2) \right \}.
\end{align*}
Analogous proof holds for the case $t_ 2 < t_1$.
\end{proof}
From these results one deduces that
\[
N_{t_1, t_2}(X_1 \times \X, \X \times X_2) \app 1 \Ra \min \left \{n_2(t_1, X_1, t_2), n(t_1, X_1, t_2,X_2) \right \} \app 1,
\]
and therefore, the consistency condition (\ref{ccr2}) is satisfied if, for all $t_1,t_2 \in T$, the following condition is satisfied:
\bg{equation} \label{lastco}
\exists P_{t_1, t_2} \in \PP: n_1(t_1, X_1, t_2), n(t_1, X_1, t_2, X_2 ) \app 1 \Ra N_{P_{t_1, t_2}}(X_1\times \X, \X \times X_2) \app 1.
\end{equation}

A further simplification consists of replacing the function $n(t, X, s, Y)$ defined by equation (\ref{mequ}) with the simpler function $N_Q(X \times \R, \R \times Y)$ defined by equation (\ref{defnq}), where, in this case, $X$ and $Y$ are measurable subsets of $\R$, and the time in the propagator $K$ defining $Q$ (equation (\ref{defk})) is set to $s-t$. A rigorous proof of condition (\ref{lastco}) with the function $n$ equal to $N_Q$ would be an important first step towards a rigorous proof of the consistency of the quantum typicality measure.

\section{Discussion and conclusion\label{disc}}

It is well known that the standard quantum formalism does not allow us to derive the joint distribution of the positions of a system of quantum particles at two or more different times because the position operators of the particles at different times (in Heisenberg representation) do not commute. In other words, the quantum formalism does not allow us to derive a probability measure on the set of trajectories of a system of quantum particles. This fact has certainly contributed to the idea that quantum particles do not follow definite trajectories, and this idea is at the base of the measurement problem of quantum mechanics.

However, the fact that the quantum formalism is not able to define a probability measure on the set of trajectories does not imply that this set does not exist, i.e., that the particles do not follow definite trajectories. Bohmian mechanics \cite{Bohm:1951,Bohm:1951xx,bmbook}, for example, rejects this conclusion and introduces a differential equation, the guiding equation, for the trajectories. The guiding equation, together with the assumption that the initial spatial distribution of the particles is that given by the wave function endows the set of trajectories with a probability measure.

The formulation of quantum mechanics proposed in this paper also assumes that the particles follow definite trajectories, but the proposed solution for the lack of a probability measure is rather different. Namely, it is assumed that the evolution of a quantum particle is not a probabilistic but rather a \tit{typicalistic} phenomenon, that is, a phenomenon that exhibits a novel kind of randomness, \tit{typicalistic randomness}. Typicalistic randomness has been presented and studied in Section \ref{typran}. Roughly speaking, the relative frequencies of the events of a typicalistic phenomenon or experiment are not stable in the long run, and therefore, it is not possible to assign them a probability. However, there is a class of events that is nearly certain, i.e., that occur nearly always in a long sequence of trials or that occurs nearly certainly in a single trial of the experiment.

The study of typicalistic randomness has required a preliminary accurate definition of the usual probabilistic randomness, and this has been the subject of Section \ref{probran}. To the best of our knowledge, although inspired by the well-known Cournot principle, the approach proposed here for defining probabilistic randomness is novel in the literature.

As a probabilistic experiment is represented by a probability space, a typicalistic experiment is represented by a \tit{typicality space}, which is basically a probability space in which the probability measure has been replaced by a much less structured \tit{typicality measure}. A typicality measure $T$ represents the almost certain events as a probability measure does, i.e., by means of the implication
\bg{equation}
T(A) \app 1 \Ra A \tx{ is nearly certain}.
\end{equation}
To be consistent, this implication requires that the typicality measure satisfies a consistency condition, namely, that there exists a probability measure $P$ such that 
\bg{equation} \label{ccc3}
T(A) \app 1 \Ra P(A) \app 1.
\end{equation}
This means for example that it is not possible that $T(A), T(A^c) \app 1$, where $A^c$ is the complement of $A$.

Returning to quantum mechanics, a way to derive a typicality measure (actually, a mutual typicality measure) from the wave function of a particle has been proposed, which is based on the notion of a separator; such a measure has been referred to as the \tit{quantum typicality measure}. Roughly speaking, a separator is (the support of) a part of the wave function that for a suitable time interval remains spatially separated from the rest of the wave function. The definition of the quantum typicality measure mathematically implements the principle that the trajectories of the particles stay nearly certainly inside the separators.

Separators are like what physicists usually call ``branches'' or ``worlds'' of the wave function, even if an unambiguous and universally accepted definition of these notions still seems lacking. In this paper, the separators are defined by an unambiguous (although vague) equation, namely, equation (\ref{nasep2}). It has been shown that, at least in the asymptotic regime, the existence of separators is a structural property of Schr\"odinger evolution, and it is not necessarily associated with the decomposition of the wave function into spatially separated wave packets, as the term ``branch'' could suggest.

With respect to the quantum typicality measure we have not been able to prove that it satisfies the consistency condition, even if some steps for simplifying the problem have been proposed. The hope is that further studies can prove (or disprove) this condition.

It is natural to compare the formulation of quantum mechanics based on the quantum typicality measure (which can be referred to as the \tit{typicalistic formulation}) with Bohmian mechanics. In fact, both formulations assume that the particles follow definite trajectories, and for both formulations the instantaneous spatial distribution of the particles is given be the Born probabilities (\ref{bornp}). Consequently the two formulations are arguably empirically equivalent because both reproduce the predictions of standard quantum mechanics and arguably predict the same quasiclassical structure of the trajectories at the macroscopic level (we recall, for example, that it is generally recognized that the Bohmian trajectories stay inside the separators (branches) of the wave function). 

We recognize that the dynamical law for the trajectories proposed by the typicalistic formulation is certainly unusual. It would already be unusual to have a probability measure rather than a differential equation as a fundamental dynamic law, but in this case we even have a measure of typicality.

On the other hand, a conceptual advantage of the typicalistic formulation with respect to Bohmian mechanics is perhaps the following. Bohmian mechanics predicts the detailed structure of the trajectories, but this structure is not empirically verifiable at the microscopic level. The predictions of the typicalistic formulation are instead much less detailed, because such a formulation says nothing about the structure of the trajectories inside the separators. In other words, the typicalistic formulation makes much less untestable predictions than Bohmian mechanics and, therefore, it is less underdetermined than Bohmian mechanics.

\bibliography{general}


\end{document}